\documentclass[twocolumn,showplaces,preprintnumbers,amsmath,amssymb,prl,aps]{revtex4-1}

\usepackage{epsfig}% Include figure files
\usepackage{graphicx}% Include figure files
\usepackage{bm}% bold math
\usepackage{float} % for pinning a figure
\usepackage{amsmath}
\usepackage{gensymb}% for degree
\usepackage{epstopdf}
\usepackage[colorlinks=true, linkcolor=blue, citecolor=blue, urlcolor=black, backref=false]{hyperref}

\begin{document}

\title{Dimensionality, nematicity and Superconductivity in Fe-based systems}

%% Notice placement of commas and superscripts and use of &
%% in the author list

\author{Khadiza Ali and Kalobaran Maiti}

%\begin{document} %for Nature documentclass

%\maketitle

%\begin{affiliations}
\altaffiliation{Corresponding author: kbmaiti@tifr.res.in}
\affiliation{Department of Condensed Matter Physics and Materials' Science, Tata Institute of Fundamental Research, Homi Bhabha Road, Colaba, Mumbai - 400005, INDIA}

%\end{affiliations}

\begin{abstract}
Study of Fe based compounds have drawn much attention due to the discovery of superconductivity as well as many other exotic electronic properties. Here, we review some of our works in these materials carried out employing density functional theory and angle resolved photoemission spectroscopy. The results presented here indicate that the dimensionality of the underlying electronic structure plays important role in deriving their interesting electronic properties. The nematicity found in most of these materials appears to be related to the magnetic long range order. We argue that the exoticity in the electronic properties are related to the subtlety in competing structural and magnetic instabilities present in these materials.
\end{abstract}

\maketitle

%\begin{document}

\section{Introduction}

Discovery of superconductivity in varied systems starting from densely populated valence band systems \cite{cuprate,Kamihara1} to very poorly populated systems \cite{Ramky_Bi} makes it one of the most outstanding challenges in contemporary condensed matter physics. While significant effort has been directed to find superconductors with elevated superconducting transition temperature ($T_c$) keeping an eye on technological applications, the origin of superconductivity is still debated even in conventional superconductors, which are believed to be well captured by the Bardeen-Cooper-Schrieffer (BCS) theory \cite{LaB6,ZrB12-1,ZrB12-2,Chainani2001,TiN,MgB2-patil}. Cuprate superconductors, Fe-based superconductors (FeSC), heavy Fermion superconductors (HFSC), {\it etc.} are different from the conventional superconductors and exhibit interesting Fermi surface reconstructions as a function of doping and/or temperature leading to the exotic phases.

\begin{figure}
 \centering
% \vspace{-18ex}
% \hspace{-6ex}
% \includegraphics[width=0.4\textwidth,natwidth=1900,natheight=1900]{Fig1-phase.jpg}
\includegraphics[scale=0.8]{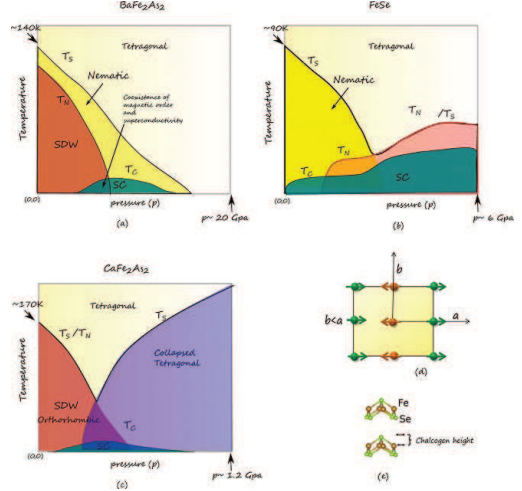}
  \vspace{-2ex}
\caption{Schematic temperature-pressure phase diagrams of (a) BaFe$_2$As$_2$ \cite{ba22}, (b)FeSe \cite{fese} and (c) CaFe$_2$As$_2$ \cite{ca22}. All these compounds undergo a structural transition from tetragonal to orthorhombic symmetry below a temperature $T_S$. In BaFe$_2$As$_2$, $T_N$ and $T_S$ are slightly different. In FeSe, the antiferromagnetic transition is absent at ambient pressure; it develops as the pressure increases. In CaFe$_2$As$_2$, $T_S$ and $T_N$ are almost the same. As the pressure increases, $T_N$ decreases gradually to zero. Eventually, $T_S$ denotes a structural change from tetragonal to collapsed tetragonal phase (cT phase) that does not show magnetic long range order. The superconductivity appears in the mixed crystallographic phase. (d) The collinear AFM order observed in Fe based compounds. (e) The anion height shown for a typical Fe-Se/As layer.}
 \label{Fig1-phase}
\end{figure}

In particular, the discovery of superconductivity in Fe-based systems \cite{Kamihara1,Kamihara2} led to a significant resurgence of interests in this field involving complex interplay of spin, charge and lattice degrees of freedom. An unique feature of these systems is the pressure induced superconductivity. The application of pressure changes the lattice constants and hence the hybridization among conduction states. Few examples of the temperature-pressure phase diagram is shown in Fig. \ref{Fig1-phase}; pressure affects the electronic phase substantially leading to varied exotic behaviors. Almost all the parent compounds exhibit paramagnetic tetragonal metallic phase at room temperature. Unlike Mott insulating ground state of parent compounds of the cuprate superconductors, the ground state of almost all the Fe-based systems are spin density wave (SDW) antiferromagnetic (AFM) metals \cite{Iron_Pnictide1,Iron_Pnictide2}. The Fermi surface of these materials is complex due to the participation of multiple $d$ bands derived from the Fe 3$d$ states in contrast to the cuprates, where single $d_{x^2-y^2}$-band plays the key role in their electronic properties.

\begin{figure}
%\vspace{-32ex}
%\hspace{-18ex}
%\includegraphics[width=0.45\textwidth,natwidth=380,natheight=700]{Fig2-gapsym.jpg}
\includegraphics[scale=0.45]{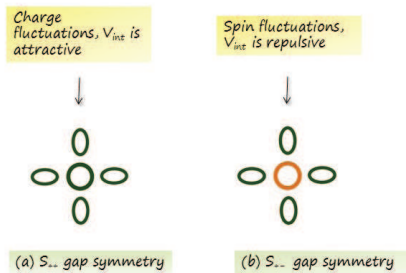}
\vspace{-12ex}
\caption{Typical superconducting gap symmetry envisaged in Fe-based systems. (a) $S_{++}$ gap symmetry. (b) $S_{+-}$ gap symmetry.}
 \label{Fig2-gapsym}
\end{figure}

The superconducting gap in cuprate superconductors possesses $d_{x^2-y^2}$ symmetry, which contains key information of the pairing mechanism. Enormous experimental and theoretical efforts led to a well accepted theory based on spin fluctuation theory \cite{cuprates_Dahmspinfluc_naturephys,thoerycuprates_Monthoux_prl1991}. It is widely accepted that superconductivity in different Fe-based materials appear due to its proximity to magnetic instabilities (spin fluctuation theory) \cite{spinfluctheory_PhysRevBScalapino1986,spinfluctheory_hirschfeld2011gapiop,
NMR_ning2010PRL,theory_naturemazin2010}. Very weak electron-phonon coupling constant in these systems led to the conclusion that the phonons may not be playing large role in the pairing of electrons for the superconductivity in the Fe and Cu-based materials \cite{Phonon_Boeri2010BaFe2As2,Phonon_PRLBoeriLaofeAs2008,Phonos_naturemonthoux2007}. The angle resolved photoemission (ARPES) data from cuprates often show signatures of strong electron-phonon coupling \cite{lanzara,zhou_prl,xie_prl} and question the role of spin fluctuation as a stand alone pairing mechanism.

ARPES results of Fe based systems are complex and the gap function is not conclusive yet. The results of inelastic neutron scattering measurements indicate sign reversed gap structure, on `1111' \cite{SpmNeutron_shamoto2010PRL}, `122' \cite{SpmNeutron_naturechristianson2008}, and `11' \cite{SpmNeutron_QiPRL2009} materials. Here, the numbers representing the material class is defined based on the number of constituent elements in their chemical formula unit; for example, $R$Fe$_2$As$_2$ ($R$ is an alkaline earth metal) is represented by `122' class. The $d$-type pairing was conclusively discarded based on these experiments. In addition, ARPES results demonstrated nodeless isotropic gap suggesting a $s$-type pairing \cite{SpmARPES_johnston2010puzzle,SpmARPES_xu2010naturephysics}. Various phase sensitive measurements suggest $s_{\pm}$ symmetry of the gap function \cite{Spm_hoffman2010signScience,Spm_Hicks2008JPSJ,Spm_sciencehanaguri2010,Spm_zeng2010naturephysics} as shown in Fig. \ref{Fig2-gapsym}. It is believed that strong antiferromagnetic spin fluctuations favour the $s_{\pm}$ type pairing. In Fe based compounds, the weakening of spin density wave state as well as the structural changes can give rise to spin fluctuations, charge fluctuations or both.

While there are differences in their electronic properties, gap symmetry {\it etc.}, all these systems exhibit similarity in the topology of the electronic structure; the effective two-dimensional behavior - here, the dimensionality is defined on the basis of the mobility of electrons in the solid. It is believed that spin fluctuations due to electron-electron Coulomb repulsion (electron correlation) and Hund's coupling is responsible for the superconductivity. It is to note here that even a BCS superconductor, MgB$_2$ having high transition temperature ($T_c$) of about 40 K also possesses effective two dimensional electronic structure.

%\section{Background}

\subsubsection*{Nematicity}

The unique feature of the coexistence of long range magnetic order and superconductivity in Fe based compounds provides an additional dimension in understanding the observed unconventional superconductivity. While the ground state of almost all the materials is a SDW state, the transition to the ground state is preceded by a structural transition from tetragonal to orthorhombic phase in most cases as shown in Fig. \ref{Fig1-phase}. In some cases, such as 122 class of materials, both the transitions occur simultaneously. Some other systems exhibit small temperature difference between the two transitions. Some systems do not show the magnetic transition (e.g. FeSe). Finding the origin of the structural phase transition is an important issue since it breaks the rotational symmetry of the parent and/or doped compounds.

The breaking of the rotational symmetry is often considered as a signature of nematicity in these materials; the emergence of nematic phase is shown in Fig. \ref{Fig1-phase}(a) and \ref{Fig1-phase}(b). The word, 'nematicity' is widely used in the field of liquid crystal to represent a type of order, where the objects forming the liquid crystals self-align to exhibit long range order; these objects can flow keeping their directional order protected although their center of mass positions are randomly distributed. In superconductors, the pairing interactions is believed to be mediated by the spin fluctuations leading to anisotropic superconducting instabilities. The nematicity is the result of the rotational symmetry breaking due to the combination of the spin fluctuations with the orbital degrees of freedom.

In Fe-based systems, two energy bands having $d_{xz}$ and $d_{yz}$ symmetries are often found to be degenerate. The structural transition might lift the degeneracy between $d_{xz}$ and $d_{yz}$ bands. The other origin of such symmetry breaking could be magnetic long range order. Numerous experimental and theoretical studies are debating on the origin of such symmetry breaking leading to nematicity in the system.

\subsubsection*{Charge density wave (CDW)}

The signature of charge density wave (CDW) state has been observed in Mn doped LaFeAsO by A. Martinelli \emph{et al}.\cite{martinelli}, which is not often discussed in Fe-based compounds.
They observed an incommensurate lattice structure in LaFeOAs for optimal Mn doping and also found a charge density wave ordering, which vanishes as the magnetism/superconductivity sets in as a function of doping.

The CDW state is a manifestation of the electron-phonon interaction in solids, which produces a modulation in charge density and can lead to structural transition. Kang \textit{et~al.}\cite{prbkang2017cdw} has shown the relevance of Peierls transition (precursor of CDW) in various structural phase transitions in La doped CaFe$_2$As$_2$.
Since the Fermi surface nesting can be related to both SDW and CDW state, in principle, both the phenomena can appear in these materials. The CDW order and superconductivity in a system usually are competing phenomena. Recent paper by Wei Li \emph{et al.} \cite{WeiLi} showed evidence of nematicity giving rise to an one dimensional charge ordering in FeSe monolayer grown on SrTiO$_3$. The large anisotropy along $a$ and $b$ axis makes the $d_{xz}$ band more itinerant than $d_{yz}$ band. This gives rise to a stripe-type charge order in the system. The Fermi surface topology changes significantly at the onset of such nematic phase although the stripe charge order exhibits temperature independence once they are formed. This observation points towards a different origin of the charge order other than the conventional Fermi surface nesting. The striped charge ordered phase appears below the nematic phase transition and expected to be originated from the magnetic fluctuations. Y. T. Tam \emph{et al.} \cite{YTTam} found that the wave vector of magnetic fluctuation ($q\simeq\pi/5$) is different from the usual nesting vector and it is comparable to the periodicity of the charge order found in FeSe films ($\approx$ 1.9~nm, where $a$ = 0.376~nm). It appears that a collinear AFM order emerges with increasing pressure in this system, while there is a charge order due to the tensile strain on SrTiO$_3$ substrate. These two different orders controlled by magnetic fluctuations are in a delicate balance even at ambient pressure.

In this article, we review some of the recent results in Fe-based systems exhibiting importance of topology in the exoticity of these materials \cite{EuFe2As2,ganesh_fetese,ganesh_CaFe2As2,kbm_confa,kbm_confb,kbm_confc,dft_khadiza}. The results were obtained by employing density functional theory (DFT) and ARPES measurements. Both theoretical and experimental results exhibit interesting scenario of complex role of the structural parameters in the electronic properties of these systems.

\section{Experimental and Theoretical details}

The electronic structure calculations were carried out following full potential linearized augmented plane wave method (FLAPW); we used Wien2k software for the calculations \cite{wien2k}. In this method, the convergence to the ground state was achieved self consistently by fixing the energy convergence criteria to 0.0001 Rydberg ($\sim$1 meV) using 10$\times$10$\times$10 $k$-points in the Brillouin zone. For the Fermi surface calculations, 39$\times$39$\times$10 $k$-points were used. We have used the Perdew-Burke-Ernzerhof generalized gradient approximation (GGA) for the density functional theoretical calculations. The Fermi surfaces were calculated using Xcrysden \cite{crysden}. In the tetragonal phase (space group $I4/mmm$), the lattice parameters are, $a$ = 3.8915 \AA, $c$ = 11.69 \AA, and $z_{As}$ = 0.372. Orthorhombic structure of CaFe$_2$As$_2$ appearing at low temperature and ambient pressure has $Fmmm$ space group with the lattice parameters $a$ = 5.506 \AA, $b$ = 5.450 \AA, $c$ = 11.664 \AA. The collapsed tetragonal (cT) phase possesses the same space group as tetragonal phase but a reduced $c$ axis and slightly increased $a$ axis, which leads to an overall reduction in cell volume. The lattice parameters for the cT phase at $P$ = 0.63 GPa are $a$ = 3.978 \AA, $c$ = 10.6073 \AA; we used $z_{As}$ = 0.372 and 0.3663 for the calculations of cT phase.

Electronic structure can be probed experimentally using photoelectron spectroscopy, which is a highly versatile technique and based on photoelectric effect. In this process, the electrons are excited from its ground states using a photon beam and determine the kinetic energy and momentum of the photoemitted electrons. The experimental spectra is directly related to the density of states/spectral functions of the system. For our experiments, high quality single crystals of CaFe$_2$As$_2$ were grown using high temperature solution growth method as described in the Refs. \cite{neraj,mittal}. The composition and structure were verified by energy dispersive analysis of $x$-rays (EDAX) followed by $x$-ray photoemission and $x$-ray Diffraction (XRD). Single crystallinity was ensured by sharp Laue Pattern. Angle resolved photoemission (ARPES) measurements were carried out at Elettra, Trieste, Italy and TIFR, Mumbai using Scienta R4000 WAL electron analyzer with an energy resolution fixed at 15 meV and angle resolution of ${\approx}$ 0.3$^{\circ}$. The samples were cleaved \textit{in~situ} along $ab$ plane yielding a mirror like clean surface. The base pressure of the spectrometer chamber was maintained at 4$\times$10$^{-11}$ Torr during cleaving and the photoemission measurements.

\section{Results and Discussions}

\subsubsection*{Fermi surface}

\begin{figure}
%\vspace{12ex}
%\hspace{-24ex}
%\includegraphics[width=0.45\textwidth,natwidth=400,natheight=380]{Fig3-FS.jpg}
\includegraphics[scale=0.45]{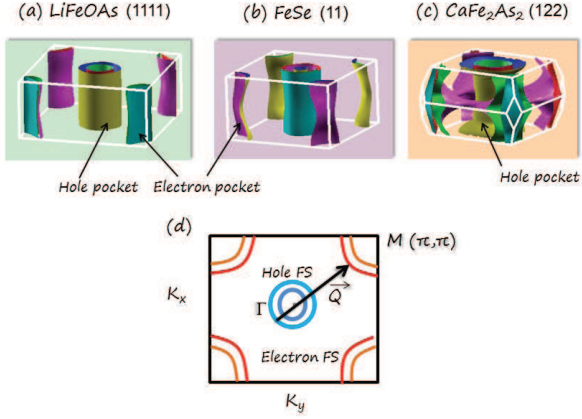}
\vspace{-2ex}
\caption{Fermi surfaces calculated using density functional theory (DFT)
for different classes of Fe based superconductors. Fermi surfaces are represented by colored surfaces for different energy bands. The results for (a) LiFeOAs, (b) FeSe and (c) CaFe$_2$As$_2$ are shown. Surfaces at the Brillouin zone center are the hole Fermi surfaces and the ones at the zone corners are electron Fermi surfaces. The cylindrical shape suggests effective two dimensional symmetry in every case. (d) A two-dimensional cut of Fermi surface showing the nesting vector $Q(\pi,\pi)$ between the hole Fermi surface (sky color) and electron Fermi surface (red).}
\label{Fig3-FS}
\end{figure}

The challenge in the Fe based superconductors is the multiband nature of the valence band. In Fig. \ref{Fig3-FS}, we show the calculated Fermi surface of few Fe-based systems, namely LiFeOAs, FeSe and CaFe$_2$As$_2$. Almost all these energy bands exhibit cylindrical shape indicating their effective two-dimensional nature as expected for such layered systems. The enclosed surfaces at the center represent the hole pockets and the ones at the zone corners are the electron pockets. The shape of the Fermi sheets are almost perfect cylinders in 1111 and 11 compounds. In 122 case (CaFe$_2$As$_2$), the cylinders are slightly distorted at the top and bottom surfaces. A typical cross section of the Fermi surfaces is shown in Fig. \ref{Fig3-FS}(d). One of the hole pockets around $\Gamma$ point is nested with an electron pocket at the zone corner; such nesting leads to the formation of SDW/CDW states in this system. Evidently, the Fe based superconductors are complex and provide a large platform to study the origin of superconductivity in high $T_c$ materials in detail, along with its possible connection to magnetism and crystal structure.

\subsubsection*{Band structure}

\begin{figure}
%\hspace{-12ex}
%\includegraphics[width=0.45\textwidth,natwidth=400,natheight=700]{Fig4-BndStr.jpg}
\includegraphics[scale=1.5]{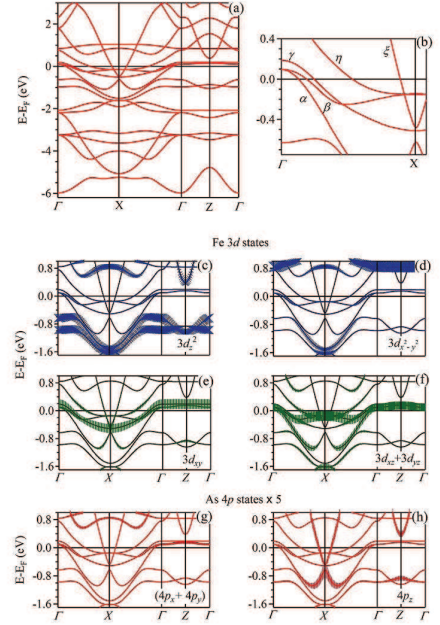}
\vspace{-2ex}
\caption{(a) Energy band structure of CaFe$_2$As$_2$ in the tetragonal phase corresponding to the room temperature structure. (b) Expanded view of the band structure near Fermi level denoted by 'zero' in the energy scale and along $\Gamma$X direction. The individual bands are denoted by $\alpha$, $\beta$ and $\gamma$ forming hole pockets around $\Gamma$-point. The orbital character of the bands are shown for (c) Fe 3$d_{z^2}$, (d) Fe 3$d_{x^2-y^2}$, (e) Fe 3$d_{xy}$, (f) Fe (3$d_{xz}$ + 3$d_{yz}$), (g) As (4$p_x$ + 4$p_y$) and (h) As 4$p_z$ electronic state contributions. The As 4$p$ contributions are enhanced by 5 times to have better visibility of their contribution close to the Fermi level.}
\label{Fig4-BndStr}
\end{figure}

We now discuss the energy band structure of CaFe$_2$As$_2$ in Fig. \ref{Fig4-BndStr}. The character of the bands are shown in the lower panels; Fig. \ref{Fig4-BndStr}(c) - \ref{Fig4-BndStr}(h). Several energy bands appear to cross the Fermi level leading to metallic behavior of the system. Three energy bands near the $\Gamma$ point, denoted by $\alpha$, $\beta$ and $\gamma$ form three hole pockets as shown in Fig. \ref{Fig4-BndStr}(b) in an expanded energy and $k$ scales. Two energy bands, $\eta$ and $\xi$ cross the Fermi level nearer to $X$-point and form electron pockets. The nesting between one of the hole pocket around $\Gamma$-point ($\beta$-band) and an electron pocket around $X$-point ($\xi$-band) leads to SDW phase in the ground state of this material; a similar scenario is observed in almost all the materials in this class.

It is evident from the figure that the Fe 3$d$ states have dominant contribution in the energy range close to the Fermi level (-1.6 eV to $\epsilon_F$), which are called antibonding bands. The bonding bands appear between -6 to -2 eV possessing major part of the As 4$p$ contributions (not shown in the figure for clarity). From the analysis of the character of the bands, it appears that the $\gamma$ band possesses mainly $d_{xy}$ character, and the $\alpha$ \& $\beta$ bands have contributions from degenerate $d_{xz}$ and $d_{yz}$ orbitals \cite{ganesh_CaFe2As2}. These three bands possess $t_{2g}$ symmetry. The $e_g$ bands constituted by $d_{x^2-y^2}$ and $d_{z^2}$ spin-orbitals appear far away from the Fermi level and hence do not play significant role in the electronic properties of the system. Overall, the bands near the Fermi level has very less As 4$p$ character except one electron pocket at $X$, which has As 4$p_z$ character. This is in line with the expectation from the fact that As layers appear above and below Fe layers and hence $p_z$ orbital is expected to play a major role in the Fe 3$d$-As 4$p$ hybridization.

\begin{figure}
%\vspace{-48ex}
%\hspace{-12ex}
%\includegraphics[width=0.4\textwidth,natwidth=380,natheight=800]{Fig5-Ca122band.jpg}
\includegraphics[scale=1.2]{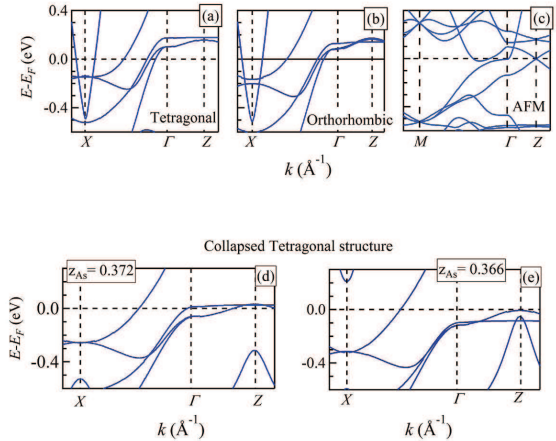}
\vspace{-2ex}
\caption{Energy band structure of CaFe$_2$As$_2$ in (a) tetragonal, (b) non-magnetic orthorhombic and (c) antiferromagnetic orthorhombic phases. While non-magnetic tetragonal and orthorhombic solutions look very similar, the onset of AFM order introduces significant spectral redistribution. The band structure for collapsed tetragonal phase for (d) $z_{As}$ = 0.372 and (e) $z_{As}$ = 0.366. Compression of the $c$-axis without change in $z_{As}$ (=0.372) leads to vanishing of two hole pockets. In the crystal structure at $P$ = 0.63 GPa, the $z_{As}$ becomes 0.366, and all the hole pockets vanishes leading to vanishing of nesting and hence, magnetic order.}
\label{Fig5-Ca122band}
\end{figure}

The influence of crystal structure without changing the magnetic phase is discussed in Fig. \ref{Fig5-Ca122band}. Here, it is clear that change in structure from tetragonal to orthorhombic symmetry does not have significant effect in the energy band structure. The Fermi pockets remain very similar, with a small compression in the low temperature phase. The magnetic order leads to significant reconstruction of the Fermi pockets. In the cT-phase, the crystal symmetry remains tetragonal and $c$-axis gets compressed. Here, we show the change in band structure in different phases along $X-\Gamma-Z$ symmetry lines. In the AFM phase, the corresponding $k$-vector is in $M-\Gamma-Z$-direction. The transition to cT phase leads to a destruction of the hole pockets. Interestingly, if we use $z_{As}$ value similar to that observed in tetragonal/orthorhombic phase and change only the lattice constants as found in the cT phase, the band structure exhibit large shift of the bands towards higher binding energies. Consequently, the hole pockets corresponding to $\alpha$ and $\beta$ bands vanishes and the one corresponding to $\gamma$ band barely survives. The results for $z_{As}$ corresponding to $P$ = 0.63 GPa pressure \cite{ARPES_cT1,ARPES_cT2} exhibit disappearance of all the hole pockets and in the process the Fermi surface nesting will also disappear leading to destruction of magnetic long range order. It is evident that one can control the presence/disappearance of the hole pockets by tuning the structural parameters, which play the key role in deriving magnetic long range order.

\subsubsection*{ARPES}

\begin{figure}
%\vspace{-12ex}
%\hspace{-12ex}
%\includegraphics[width=0.45\textwidth,natwidth=300,natheight=700]{Fig6-FigFS.jpg}
\includegraphics[scale=1.5]{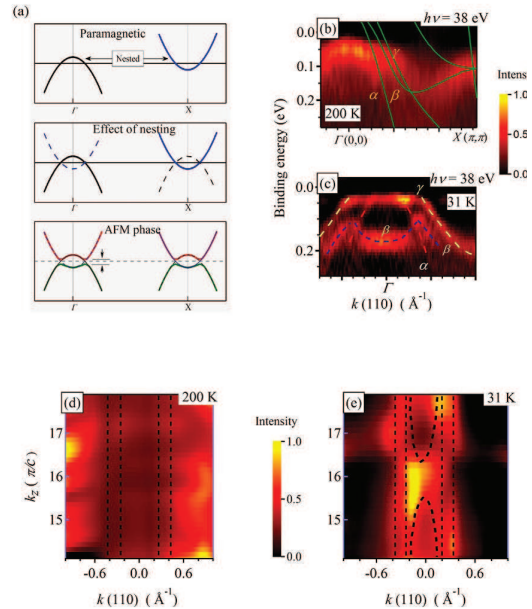}
%\vspace{-2ex}
\caption{Angle resolved photoemission spectroscopic (ARPES) results of CaFe$_2$As$_2$. (a) Schematic description of the gap formation arising from magnetic order related to Fermi surface nesting. (b) Second derivative of the experimental data at 200 K obtained using photon energy of 38 eV. The lines are the calculated results - here, the energy scale of the calculated results are compressed by 40\%. (c) Second derivative of the ARPES data at 31 K and $h\nu$ = 38 eV. A clear signature of the folding of $\beta$ band due to AFM order is observed. The Fermi surfaces in the $k_xk_z$ plane obtained by varying incident photon energies. (d) The results at 200 K exhibit two-dimensional geometry. (e) The results at 31 K show two dimensionality for two bands and an additional one having $k_z$ dependence. Dashed lines are the guide to the eye exhibiting the Fermi surface topology.}
\label{Fig6-FigFS}
\end{figure}

In Fig. \ref{Fig6-FigFS}(a), demonstrate how SDW state influences the energy band structure. The energy bands are nested along $\Gamma X$ direction. At the onset of magnetic order, the replica of the band at $\Gamma$ will appear at $X$ and vice versa, which will lead to an opening of a band gap as shown schematically. In Fig. \ref{Fig6-FigFS}(b), we show the experimental ARPES results collected at 200 K using photon energy of 38 eV, which is close to the $k_z$ value of 14$\pi$/$c$ and corresponds to $Z$ point in the $k_z$ axis. The lines superimposed on the experimental results are the calculated band structure results - we needed to compress the energy scale by 40\% to capture the experimental bandwidth \cite{Ca122-unpublished}. This suggests that the electron-electron Coulomb repulsion strength among the corresponding conduction electros is significant, which is often manifested as band narrowing due to the effective reduction of the itineracy of these electrons.

The experimental results in the magnetically ordered phase are shown in Fig. \ref{Fig6-FigFS}(c). Here, we observe two distinct bands crossing the Fermi level; these bands are marked as $\alpha$ and $\gamma$. The $\beta$ band exhibit a folding due to magnetic transition; the hole pocket corresponding to the $\beta$ band was nested with one of the electron pocket at $X$-point as demonstrated in the schematic. Below the magnetic transition temperature, these bands fold back forming an energy gap at the Fermi level, which stabilizes the magnetically ordered phase via lowering the total energy of the system. The energy gap appears to be of the order of 0.2 eV.

The experimental Fermi surfaces at 200 K and 31 K are shown in Fig. \ref{Fig6-FigFS}(d) and \ref{Fig6-FigFS}(e), respectively. While the topology of the Fermi surfaces at 200 K is cylindrical consistent with the calculated results (see Fig. \ref{Fig3-FS}), the emergence of $k_z$ dependence of the central hole pocket around the $k_z$ axis is curious and might be related to some hidden phase in the system. This possibility is conjectured as this band is additional to maximum two Fermi surfaces expected; the third one among the three bands around $\Gamma$-point already formed a gap due to magnetic long range order.

\section{Discussion}

It is clear from the above results that although the materials are three dimensional systems, the effective electronic structure is two dimensional; this is manifested in both theoretical calculations and experimental results for both tetragonal and orthorhombic symmetries. $d_{xz}$ and $d_{yz}$ bands are found to be degenerate; which is expected in tetragonal symmetry. The distortion leading to the orthorhombic phase at low temperatures is very small ($<$ 1\%) and the degeneracy lifting of these two bands are not significant even in orthorhombic phase. The neutron diffraction measurements exhibit antiferromagnetic order along $a$, while the moments are aligned in parallel along $b$-axis. The similarity between $a$ and $b$ axis is broken by the magnetic order and the degeneracy of the $d_{xz}$ and $d_{yz}$ bands are lifted further when the magnetic long range order sets in.

Before going into the possible role of orbital and spin degrees of freedom in superconductivity, we consider the issue of electron nematic order found in FeSCs in general. The nematicity spontaneously breaks the symmetry between $x$ and $y$ axis, although the underlying crystal symmetry may be tetragonal. Due to this, the observed bulk properties like resistivity, magnetization, {\it etc}. \cite{lu2014nematic_science,kasahara2012electronic_nature} exhibit dissimilarity along $x$ and $y$ directions giving rise to an effective lower symmetry than the tetragonal crystal lattice; the four fold symmetry of the tetragonal phase is reduced to a two fold symmetry with respect to the $z$ axis. It is to note here that using strain as a control parameter for resistivity measurements, Jiun-Haw Chu {\emph et al.} \cite{science12} showed divergent behavior of the nematic susceptibility and attributed the scenario to purely electronic interactions; the structural deformation merely follows the nematic order. The nematic susceptibility can be defined as $\delta \rho_{aniso}/\delta h$, where $\rho_{aniso}=(\rho_b -\rho_a)/(\rho_b+\rho_a)$ and $h$ is the strain. Signature of such anisotropy is also observed in the inelastic neutron diffraction studies showing exceptionally different nearest neighbour exchange coupling along $a$ and $b$ directions, although the lattice constant along $a$ and $b$ are minutely different \cite{diallo2009itinerant_PRL,neutron_Zhao2009NatPhys}.

The strong electron nematicity induces unidirectional electronic nanostructures in under doped Ca(Fe$_{1-x}$Co$_x$)$_2$As$_2$ ($x\sim$ 0.3) along $a$ axis \cite{chuang2010nematic_Science}. The nanostructures have dimension of $\sim$ 8$a_{Fe-Fe}$, which gives a band folding in the reciprocal space with wavevector $\simeq 2\pi/8a_{Fe-Fe}$ seen through quasiparticle interference (QPI) imaging. R. Daou \emph{et al.} has pointed out a large in-plane anisotropy of the Nerst effect in the pseudogap phase of untwined YBa$_2$Cu$_3$O$_y$ (YBCO)\cite{Daou_BRS2010Natue}, where the fourfold symmetry breaking is attributed to the onset of electron nematic order and not due to the lattice structure. The presence of various exotic phenomenon in NaFeAs and absence in similarly structured material, LiFeAs, indicate strongly material dependent behaviors \cite{davis2014iron_naturephysics,NAFeAs2014NatPhys,NaFeAs_DengPhysRevB}. The independence of resistivity anisotropy on disorder and impurity studied extensively on Co and Ni doped BaFe$_2$As$_2$ showed the robustness of nematic order \cite{robustKuo_PhysRevLet}.

\begin{figure}
% \centering
% \vspace{2ex}
% \hspace{-10ex}
% \includegraphics[width=0.45\textwidth,natwidth=1200,natheight=900]{Fig7-nemat.jpg}
 \includegraphics[scale=0.4]{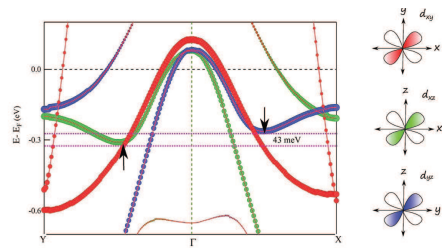}
  \vspace{-0ex}
\caption{Calculated band structure in orthorhombic phase without considering the magnetic order. The energy dispersion is shown along two high symmetry lines that are equivalent in tetragonal phase but becomes non equivalent in orthorhombic phase due to lattice distortion. The colour and thickness of the energy band denotes the orbital character and their weight. The degeneracy of $d_{xz}$ and $d_{yz}$ bands seems to be lifted due to structural transition.}
\label{Fig7-nemat}
\end{figure}

All these phenomena in diverse systems indicate that nematicity is an intrinsic behavior of these materials and it is difficult to attribute this to a particular control parameter. In the tetragonal phase of CaFe$_2$As$_2$, the orbitals $d_{xz}$ and $d_{yz}$ follow the underlying lattice symmetry and are degenerate. In the orthorhombic phase below 170 K, $x$ and $y$ directions become non equivalent due to the distortion and lifts the degeneracy between $d_{xz}$ and $d_{yz}$. In Fig. \ref{Fig7-nemat}, we have shown the band dispersion along $x$ and $y$ directions in the reciprocal space, which were equivalent in the tetragonal structure. We observe an energy separation of about 43 meV between the bottom of the two bands; the dispersion of $d_{yz}$ along $\Gamma X$ is weaker by 43 meV than the dispersion of $d_{xz}$ along $\Gamma Y$. It is to note here that the calculated splitting in BaFe$_2$As$_2$ is very small ($\sim$10 meV), whereas ARPES experiment show a large splitting of 80 meV even above the structural phase transition in detwinned single crystals \cite{pnas}. Evidently, the calculated results are not adequate to capture the experimental observations in BaFe$_2$As$_2$; the same needs to be investigated in CaFe$_2$As$_2$.

The scenario observed in the calculated results of CaFe$_2$As$_2$ may be along expected line as the distortion in lattice parameters is too tiny (tetragonal to orthorhombic, $<1\%$) to have a prominent signature of electronic nematicity \cite{science12,kuo}. Moreover, the signature of the change in electronic structure appears well above the structural phase transition, which is closely followed by a collinear antiferromagnetic order in `1111' and `122' systems; a precursor effect-type behavior \cite{bairo3,ca3co2o6,sampath}. Thus, one needs to go beyond the structural changes in these systems.

In the ARPES results of CaFe$_2$As$_2$, the Fermi surface mapping shown in Fig. \ref{Fig6-FigFS} shows signature of two dimensional topology of the energy bands associated to tetragonal and orthorhombic phase - this can be ascertained by counting the number of energy bands to be three ($d_{xy}$, $d_{xz}$ and $d_{yz}$). The additional band exhibiting three dimensional symmetry appears to be related to some hidden order; more studies are required to understand the origin of such bands.

It is important to note here that $\alpha$ and $\beta$ bands possessing ($d_{xz}$+$d_{yz}$) symmetry shown in Fig. \ref{Fig5-Ca122band} exhibit dispersion along $\Gamma Z$ direction as expected due to the finite hybridization of these states with As $p_z$ electronic states and Ca states are also significantly hybridized with the As 4$p$ states \cite{dft_khadiza}. The $\gamma$ band is flat along $\Gamma Z$ direction exhibiting almost perfect two-dimensional symmetry. Since, all these bands cross the Fermi level and the energy eigenstates exhibiting three dimensionality essentially is a part of the unoccupied part of the electronic structure far away from the Fermi level, this is not captured in the photoemission spectra representing the occupied part of the electronic structure. In the cT-phase, the compression along $c$-direction and $z_{As}$ leads to a shift of these bands towards higher binding energies; the $\xi$-band becomes completely unoccupied that keeps the total electron count conserved. This change brings the degenerate $\alpha$ and $\beta$ band in proximity of the Fermi level near $Z$-point while they are far away from the Fermi level at $\Gamma$-point exhibiting significant $k_z$-dependence.

\section{Conclusions}

In summary, we reviewed the electronic structure of some of the Fe-based systems exhibiting superconductivity on application of pressure and/or doping of charge carriers. Almost all these materials exhibit magnetic and structural transition leading to a magnetically ordered ground state. While the structural transition involves small distortion of the crystal lattice leading to a rotational symmetry breaking, the effect in the electronic structure appears to be much less than the observed nematicity via experiments. One the other hand, magnetic long range order influences the lifting of the degeneracy of the $d_{xz}$ and $d_{yz}$ bands more strongly, which can be associated to the observed strong nematicity of the system. All these results lead to the conclusion that a major driving force of the nematicity in Fe-based systems is electronic in nature and magnetic order plays important role in this.

\section{Acknowledgements}

K. M. acknowledges financial assistance from the Department of Science and Technology, Govt. of India under J.C. Bose Fellowship program and Department of Atomic Energy, Govt. of India under DAE-SRC-OI award scheme.

%\section{References}

%%%%%%%%%%%%%%%%%%%%%%%%%%%%%%%%%%%%%%%%%%%%

%%%%%%%%%%%%%%%%%%%%%%%%%%%%%%%%%%%%%%%%%%%%

%%%%%%%%%%%%%%%%%%%%%%%%%%%%%%%%%%%%%%%%%%%%

%%%%%%%%%%%%%%%%%%%%%%%%%%%%%%%%%%%%%%%%%%%%

\end{document}